\documentclass[preprint]{emulateapj}

\newcommand{\mjybeam}{mJy beam$^{-1}$}

\shorttitle{The JCMT Debris Disk Survey}
\shortauthors{Matthews et al.}
\begin{document}

\title{An Unbiased Survey of 500 Nearby Stars for Debris Disks: A
  JCMT Legacy Program}

\author{Brenda C.\ Matthews\altaffilmark{1},
Jane S.\ Greaves\altaffilmark{2},
Wayne S.\ Holland\altaffilmark{3},
Mark C.~Wyatt\altaffilmark{4},
Michael J.\ Barlow\altaffilmark{5},
Pierre Bastien\altaffilmark{6},
Chas A.\ Beichman\altaffilmark{7},
Andrew Biggs\altaffilmark{3},
Harold M.\ Butner\altaffilmark{8},
William R.F.\ Dent\altaffilmark{3},
James Di Francesco\altaffilmark{1},
Carsten Dominik\altaffilmark{9},
Laura Fissel\altaffilmark{10},
Per Friberg\altaffilmark{11},
A.G.\ Gibb\altaffilmark{12}
Mark Halpern\altaffilmark{12},
R.\,J.\ Ivison\altaffilmark{3,13},
Ray Jayawardhana\altaffilmark{10},
Tim Jenness\altaffilmark{11},
Doug Johnstone\altaffilmark{1},
JJ Kavelaars\altaffilmark{1},
Jonathon L.\ Marshall\altaffilmark{14},
Neil Phillips\altaffilmark{3,13},
Gerald Schieven\altaffilmark{11},
Ignas A.G.\ Snellen\altaffilmark{15},
Helen J.\ Walker\altaffilmark{5},
Derek Ward-Thompson\altaffilmark{16},
Bernd Weferling\altaffilmark{11},
Glenn J.\ White\altaffilmark{14,17},
Jeremy Yates\altaffilmark{5},
Ming Zhu\altaffilmark{11}}

\altaffiltext{1}{Herzberg Institute of Astrophysics, National Research Council
  of Canada, 5071 West Saanich Road, Victoria, BC, V9E 2E7 Canada}
\altaffiltext{2}{School of Physics and Astronomy, University of St Andrews,
  North Haugh, St Andrews, Fife KY16 9SS, UK}
\altaffiltext{3}{UK Astronomy Technology Center, Royal Observatory, Blackford Hill, Edinburgh EH9 3HJ, UK}
\altaffiltext{4}{Institute of Astronomy, University of Cambridge, Madingley
  Road, Cambridge, CB3 0HA, UK}
\altaffiltext{5}{University College London, Gower Street, London, WC1E
  6BT, UK}
\altaffiltext{6}{D\'{e}partement de physique et Observatoire du Mont-M\'{e}gantic, Universit\'{e} de Montr\'{e}al, C. P. 6128, Succ. Centre-ville, Montr\'{e}al, QC H3C 3J7, Canada}
\altaffiltext{7}{Michelson Science Center, California Institute of
  Technology, 770 South Wilson Avenue, Pasadena, CA 91125, U.S.A}
\altaffiltext{8}{Department of Physics and Astronomy, James Madison
  University, Harrisonburg, VA 22807, U.S.A}
\altaffiltext{9}{Astronomical Institute ``Anton Pannekoek,'' University
  of Amsterdam, Kruislaan 403 NL-1098 SJ Amsterdam, The Netherlands}
\altaffiltext{10}{Department of Astronomy and Astrophysics, University of Toronto, 50 St. George Street, Toronto, ON M5S 3H4, Canada}
\altaffiltext{11}{Joint Astronomy Centre, Hilo, HI}
\altaffiltext{12}{Department of Physics \& Astronomy, University of British Columbia, Vancouver, BC V6T 1Z1, Canada}
\altaffiltext{13}{Scottish Universities Physics Alliance, Institute for Astronomy, University of
Edinburgh, Royal Observatory, Blackford Hill, Edinburgh EH9 3HJ, UK}
\altaffiltext{14}{Department of Physics and Astronomy, The Open
  University, Milton Keynes MK7 6AA, U.K.}
\altaffiltext{15}{Leiden Observatory, Leiden University, Postbus 9513, 2300 RA, Leiden, the Netherlands}
\altaffiltext{16}{School of Physics and Astronomy, Cardiff University,
  Queens Buildings, The Parade, Cardiff CF24 3AA, U.K.}
\altaffiltext{17}{Rutherford Laboratory, Space Science \& Technology
  Department, CCLRC Rutherford Appleton Laboratory, Chilton, Didcot,
  Oxfordshire OX11 0QX, U.K.}

\email{brenda.matthews@nrc-cnrc.gc.ca}

\begin{abstract}
We present the scientific motivation and observing plan for an
upcoming detection survey for debris disks using the James Clerk
Maxwell Telescope.  The SCUBA-2 Unbiased Nearby Stars (SUNS) survey
will observe 500 nearby main sequence and sub-giant stars (100 of each
of the A, F, G, K and M spectral classes) to the 850 \micron\
extragalactic confusion limit to search for evidence of submillimeter
excess, an indication of circumstellar material.  The survey distance
boundaries are 8.6, 16.5, 22, 25 and 45 pc for M, K, G, F and A stars,
respectively, and all targets lie between the declinations of
$-40^\circ$ to $80^\circ$. In this survey, no star will be rejected
based on its inherent properties: binarity, presence of planetary
companions, spectral type or age.  The survey will commence in late
2007 and will be executed over 390 hours, reaching 90\% completion
within two years.  This will be the first unbiased survey for debris
disks since the {\it InfraRed Astronomical Satellite}.  We expect to
detect $\sim 125$ debris disks, including $\sim 50$ cold disks not
detectable in current shorter wavelength surveys.
To fully exploit the order of magnitude increase in debris disks
detected in the submillimeter, a substantial amount of complementary
data will be required, especially at shorter wavelengths to constrain
the temperatures and masses of discovered disks.  High resolution
studies will likely be required to resolve many of the disks.
Therefore, these
systems will be the focus of future observational studies using
a variety of observatories, including Herschel, ALMA, and JWST, to
characterise their physical properties.  For non-detected systems,
this survey will set constraints (upper limits) on the amount of
circumstellar dust, of typically 200 times the Kuiper Belt mass, but
as low as 10 times the Kuiper Belt mass for the nearest stars in the
sample ($\approx 2$ pc).

\end{abstract}

\keywords{stars: circumstellar matter --- submillimeter, stars:
  circumstellar disks}

\section{Introduction}

Debris disks are the dust disks found around many nearby main sequence
stars.  The dust is short-lived and so must be continuously
replenished by the destruction of comets and asteroids in these
systems, deduced to lie in fairly narrow belts between 10 to 200 AU
from the host stars.  The {\it InfraRed Astronomical Satellite (IRAS)}
was the first and only large unbiased survey of debris disks, showing
that they occur around $\sim 15$\% of nearby stars
\citep{bp93,pv99,aum84}.  There is evidence of a substantial
population of disks too cold to have been detected by IRAS and which
are only accessible to submillimeter observations \citep{les06,wdg03}
surrounding a 5-15\% of stars.  This is substantiated by the
results of \cite{rhe07} who note that $\sim 10$\% of stars show 70
\micron\ dust emission in Spitzer MIPS maps, but below IRAS
sensitivity (either due to low mass or cold dust), suggesting the
overall frequency of disks may be as high as 25\%.  Based on an
assumed disk incidence of 25\%, our survey of 500 stars could yield
as many as 125 disk detections, of which $\approx 50$ could be
cold disks.  The survey is of sufficient size to measure the disk
frequency across spectral type, and as a function of age and
multiplicity.  

Submillimeter observations have already been pivotal in studies of
debris disks, having imaged or discovered seven of the fourteen
resolved disks and constrained the
mass and temperature and hence radial extent of many of the
remainder.  The submillimeter emission is both optically-thin and
sensitive to relatively large, cold grains which dominate the massses
of these disks. Hence, our survey will provide immediate mass
estimates for detected disks. Constraining the temperature of newly
identified disks will require complementary data in the mid- or far-IR.  
This will be particularly critical for warm disks, since the two
potential submillimeter wavelengths observable through our survey (850
and 450 \micron) will both lie longward of the peak of the spectral
energy distribution (SED) and provide little constraint on the
temperature. 

The study of these disks is revolutionizing our understanding of
planet formation.  For the fourteen disks which have been resolved,
observed structures have even been used to infer the location of
unseen planets \citep[e.g.,][]{wya03}.  Many more disks have been
characterized by their SEDs, showing that they are the extrasolar
equivalents of the Kuiper and asteroid belts of the Solar System.  The
radial positions and masses of these belts, particularly when this
information can be compared for stars of different ages, spectral
types, multiplicity or known planetary companions, provide vital
constraints on planet formation processes and on how the resulting
planetary systems subsequently evolve.

SCUBA-2 \citep{hol06} is a new submillimeter camera arriving at the
James Clerk Maxwell Telescope (JCMT) in late 2007. SCUBA-2 is expected
to be, per pixel, five times as sensitive as its predecessor SCUBA
\citep[Submillimeter Common User Bolometer Array,][]{hol99} at 850
\micron.  Its larger (7\arcmin\ $\times$ 7\arcmin), fully-sampled
field of view makes it a premiere survey instrument, and seven
comprehensive surveys have been planned to maximize the scientific
output from the JCMT over the next five years (Ward-Thompson et
al. 2007, in prep; SLS paper \citep{plu07}.  Like its predecessor, it
observes simultaneously at 850 and 450 \micron, with resolutions of
15 and 7\arcsec, respectively. Unlike its predecessor, SCUBA-2 will
Nyquist sample the sky at 850 \micron; at 450 \micron, the array is
undersampled by a factor of two.  The focal plane area coverage is
provided by four ``sub-arrays'' at each wavelength, where each
sub-array is comprised of $32 \times 40$ detectors
\citep[see][]{hol06}.

The SCUBA-2 Unbiased Nearby Stars (SUNS) survey will utilize 390 hours
to observe 500 nearby stars (the 100 nearest M, K, G, F and A stars)
to the JCMT's extragalactic confusion limit at 850 \micron\ to detect
and map circumstellar dust.  The survey is completely unbiased; no
star will be rejected due to its intrinsic properties.  The survey
will determine the incidence of disks around nearby stars, constrain
masses (and temperatures of disks detected in the far-IR), discover
disks too cold to be detected in the far-IR, and provide limits on the
presence of dust which are vital to targeted planet search missions
such as Darwin and the Terrestrial Planet Finder (TPF), as well as
future missions which will resolve disks in unprecedented detail,
i.e., the Atacama Large Milliemeter-submillimeter Array, the James
Webb Space Telescope, and the Far-IR Interferometer
\citep[FIRI,][]{ivi05}.

The mass sensitivity of the survey is a strong function of both
distance and disk temperature.  For disks of 70 K, the sensitivity
ranges from $0.003 - 1.4 M_{\rm lunar}$ ($M_{\rm lunar} = 1/81 \
M_\oplus$) for the nearest (2 pc) and furthest (45 pc) stars. For
lower disk temperatures of 40 K, the range is $0.005 - 2.7 M_{\rm
lunar}$.  At the mean distance of the survey stars (15 pc), we will be
sensitive to dust masses typically $\sim 200$ times the dust mass of
the Kuiper Belt ($\approx 10^{-5} M_\oplus$), which is the mass of the
disk around $\epsilon$ Eridani.  Thus, while it will not be sensitive
to present day Solar System analogues, it will detect such systems
which are in a period of unusually high dust mass.  We know that the
Kuiper Belt is a factor of $\approx 100$ times less massive than
expected from the distribution of solids in the Solar System
\citep{mor04}.  Thus the Kuiper Belt used to be more massive.
Equally, an episodic mode of dust creation could render the Kuiper
Belt detectable in the future. The mass limit for undetected systems
will also be useful for future planet search missions.  Missions such
as Darwin/TPF require a dust free system, typically below 10 times
that of the Solar System, to limit the integration time required to
detect Earth-like planets \citep{bei04}.

The SUNS survey will provide a significant legacy for the field of
extrasolar planetary system research. This field is rapidly evolving, and
a variety of techniques are being developed to characterize the
planetary systems of nearby stars. This survey will determine the dust
and planetesimal content of these systems and so provide vital
complementary information on the outcome of planet formation in them;
for some techniques, the dust content even provides the limiting
factor determining whether such techniques are going to work \citep[e.g.,
TPF,][]{bei06b}.

In this paper, we provide details about the SUNS survey, including the
motivation for the survey in the context of our current understanding
of debris disk systems ($\S$ \ref{motivation}), advantages of the
submillimeter compared to shorter wavelengths ($\S$ \ref{adsubmm}),
and the mass sensitivity of our survey ($\S$ \ref{masssen}).
Our science goals are described in $\S$ \ref{science},
and we describe the details of our target list in $\S$ \ref{targets},
including ancillary targets, subset populations and statistics.  Plans
for complementary data to the SUNS survey are described in $\S$
\ref{complementary}.  We describe the data products in $\S$
\ref{dataproducts} and summarize the survey in $\S$ \ref{summary}.

\section{Motivation for the Survey}
\label{motivation}

\subsection{Current Status of Debris Disk Studies}

All of the approximately 200 known candidate debris disks were first
discovered by their thermal emission, which is brighter than the
photospheric emission of their host stars at far-IR and longer
wavelengths \citep[e.g.,][]{man98}.  The majority of these disks and
disk candidates were discovered by {\it IRAS}, which provided the
first and only large unbiased survey of nearby stars for excess
thermal emission at 12-100 \micron\ (with resolutions 0.5\arcmin\ to
2\arcmin). This survey showed that $\sim 15$\% of nearby stars exhibit
detectable excess emission \citep{bp93}, and the {\it IRAS} disk
candidates have been the subject of intense follow-up observations
from the ground at a range of wavelengths from optical to millimeter.
Re-analysis of the IRAS and ISO databases have resulted in additions,
and notably revisions, to the stars with identified infrared excess;
for instance, \cite{rhe07} found 153 IRAS excess stars among Hipparcos
dwarfs, 37 of which are newly identified.  Forty-eight of their excess
stars are among the 60 disk candidates identified by \cite{moo06}
through a re-analysis of {\it IRAS} and {\it ISO} targets, eleven of
which were previously unknown.  Based on these analyses, Rhee et al.\
revised the fraction of nearby A stars with detected 60 \micron\
excess to 20\%. 

Submillimeter observations have been pivotal in follow-up studies of
these disks: imaging with SCUBA had mapped or discovered seven of the
ten resolved disks at the time of its decommissioning in
mid-2005\citep[e.g.,][]{hol98}.  Such images have also ruled out
``disk candidate'' stars which turned out to have nearby, unassociated
background sources which fell within the relatively large {\it IRAS}
beamsizes \citep[e.g.,][]{jay02}. Finally, photometric submillimeter
observations of many disks have provided the best contraints on disk
masses, as well as constraining the SED and therefore the temperature
and radial extent of the disks. \citep[e.g.,][]{she04}.

Subsequent surveys have been more modest than {\it IRAS} in the number
of stars surveyed but have probed more deeply in sensitivity. For
example, both the {\it ISO} and {\it Spitzer} strategies targeted
several well-defined samples, each of which comprised at most 100
stars. The emerging picture is that the fraction of stars with
detectable disks is a function of both stellar age
\citep{rhe07,spa01}, spectral type \citep{hab01}, and wavelength
\citep{lau02}.  \cite{wer06} summarized the early results from Spitzer
surveys and targeted observations, including two Legacy Surveys ``The
Formation and Evolution of Planetary Systems''
\citep[FEPS,][]{mey04,kim05,sta05,hin06} and ``Galactic Legacy
Infrared Mid-Plane Survey Extraordinaire" \citep[GLIMPSE,][]{uzp05}.
Surveys have also targeted stellar associations, such as the nearby
$\approx 10$ Myr old TW Hydrae Association \citep{uch04,low05}, Cep OB2
clusters \citep{sic06}, Orion OB1a and OB1b \citep{her06}, and the
open clusters M47 \citep{gor04} and the Pleiades \citep{gor06,sta05}.
Projects to target nearby stars have concentrated on A stars
\citep{rie05} and field FGK stars \citep{bry06,bei06a}, including
selection by known planets \citep{bei05}, metallicity \citep{bei06b},
IRAS 60 \micron\ excess \citep{che06} and age \citep{smi06}.
\cite{bei06b} concluded that there is no increased incidence of disks
around stars of increased metallicity, in agreement with earlier
results based on disks detected by IRAS \citep{gre06}.

Evidence of a disk has been detected around the white dwarf at the 
center of the Helix Nebula \citep{su07}.
{\it Spitzer} has also produced serendipitous detections of disks
around individual stars, including HD 46190 \citep{slo04} and Be stars
in the LMC \citep{kas06}.  Targeted observations toward the
well-studied debris disks around Fomalhaut \citep{sta05} and Vega
\citep{su05} have also been undertaken.

In general, few extensive ground-based searches have been made for
excess emission. Recently, however, submillimeter surveys unbiased by
previous far-IR detections have shown that there is a substantial
population of disks only accessible in the submillimeter, since they
are too cold to have been detected by {\it IRAS} \citep{les06,wdg03}.
As Table \ref{surveys} shows, while individual submillimeter surveys are typically
small (10-20 stars), the detection rate in each is 5-25\%, indicating
that this population is both real and as numerous as the disks which
were detected by {\it IRAS}.  Statistics are currently too poor to
ascertain whether the presence of such cold disks is favored around
young/old/early-type/late-type stars.   \cite{les06} derived a
detection rate of $13_{-8}^{+6}$ \% for M dwarfs between 20-200 Myr
based on the combined results of their survey and that of
\cite{liu04}. 

\begin{deluxetable*}{llcccc}[h!]
\tablecolumns{6} 
\tablewidth{0pc} 
\tablecaption{Submillimeter Debris Disk Surveys} 
\tablehead{\colhead{Reference} & \colhead{Targets} &
\colhead{$N_{\rm star}$} & \colhead{$N_{\rm disks}$} & \colhead{\%} &
\colhead{850 \micron\ 1 $\sigma$ rms [mJy]}} 
\startdata 
Wyatt et al.\ (2003) & Lindroos binaries & 22 & 3 & 13 & 1.2 \\ 
Holmes et al.\ (2003) & Nearby bright stars & 11 & 1 & 9 & 7 \\ 
Liu et al.\ (2004) & $\beta$ Pic comoving groups & 8 & 2 & 25 & 2 \\ 
Carpenter et al.\ (2004) & FEPS nearby stars & 127 & 4 & 3 & 28 \\ 
Greaves et al.\ (2005) & Nearby G Stars & 13 & 2 & 15 & 1.5 \\ 
Lestrade et al.\ (2006) & M Dwarfs & 32 & 2 & 6 & 0.7 \\
\enddata
\tablecomments{Flux limits at 850 \micron\ were adjusted from other
wavelengths through scaling of the fiducial 850 \micron\ opacity of
1.7 cm$^2$ g$^{-1}$ and the scaling relations $\kappa \propto
(1/\lambda)$ and $F_\nu \propto \nu^{(2+\beta)}$ where $\beta = 0.7$.}
\label{surveys}
\end{deluxetable*}

\begin{figure}[h!]
\vspace*{7cm}
\includegraphics{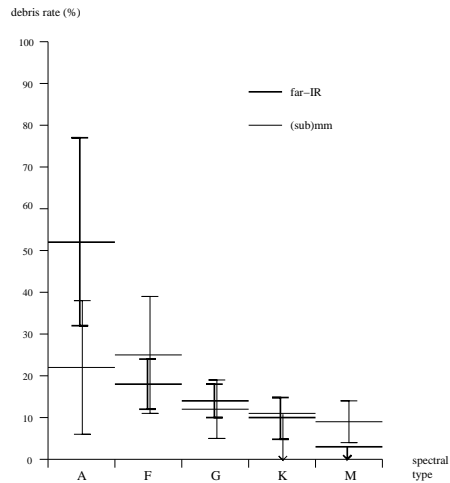}
\caption{The frequency of debris disks as a
  function of spectral type based on existing data. The errorbars are
  presently large for several spectral types. It is also evident that
  the submillimeter is the best wavelength range for which to search
  for disks around later-type stars. This is consistent with the
  detection of cold disks. The references for the data are
  \cite[Greaves et al., in prep, ][]{gau07,su06,bei06a,bry06,les06,bei05,naj05,liu04,wdg03,hol03}.}
\label{DDSincidence}
\end{figure}

The current debris disk detection rates are shown in Figure
\ref{DDSincidence}, comparing the detection rates in the far-IR to
those in the submillimeter. The uncertainties in the disk frequencies
are still large for several spectral types.  Already, it is evident 
that the submillimeter is the favorable regime for detection of disks
around low-mass stars. 

\subsection{Advantages of the Submillimeter to the Far-IR}
\label{adsubmm}

\begin{figure}[h!]
\vspace*{8cm}
\includegraphics{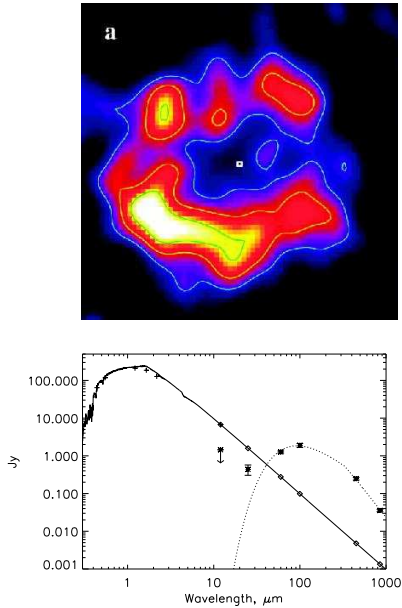}
\caption{850 \micron\ image of $\epsilon$ Eridani
  \citep{gre05} and the SEDs of the star (solid line) and grey body
  fit to the dust (dotted line).  Note the much greater disk to star
  contrast in the submillimeter compared to 60 \micron, even for this
  mid-range temperature of 55 K.}
\label{epsiloneri}
\end{figure}

\begin{figure}[h!]
\vspace*{6.5cm}
\includegraphics{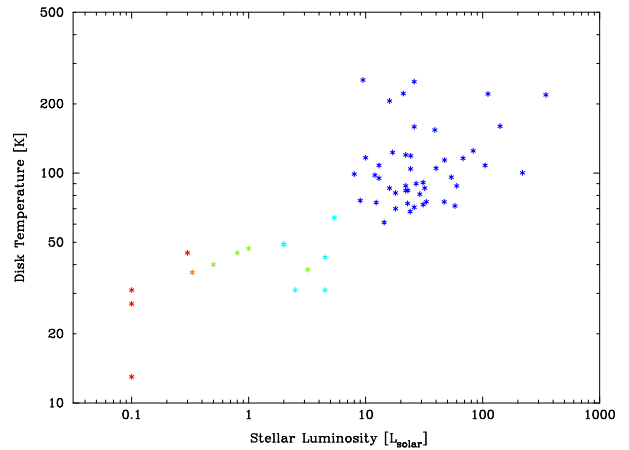}
\caption{The observed temperatures of debris
  disks versus stellar luminosities. The A stars (dark blue) are a
  sample of 46 detected at both 24 and 70 \micron\ \citep{wya07}. The
  data for the F (light blue), G (green), K (orange) and M stars (red)
  are taken from \cite[][and references therein]{les06}, with the
  addition of TWA 7 from \cite{mat07}.}
\label{lum.v.T}
\end{figure}

To understand the origin and frequency of debris disks, we require an
unbiased survey.  In the past, large surveys for disks have been
conducted at near-IR wavelengths \citep[e.g.,][]{mam02,hai01};
however, these only sample warm dust in the inner $\approx 1$ AU. In
the far-IR, major surveys with {\it IRAS}, {\it ISO} and {\it Spitzer}
\citep[e.g.,][and references therein]{spa01,dec03,wer06} sample the
dust at $\approx$ 50-100 K (typically 40--100 AU radius), so these are
also biased towards warm dust disks. Yet we know cool debris disks do
exist; Figure \ref{epsiloneri} shows the SED of the bright nearby disk
around $\epsilon$ Eridani, and Figure \ref{lum.v.T} shows the fitted
temperatures of known debris disks versus stellar luminosity for stars
of differing spectral classes.  Disks around lower mass stars tend to
be significantly cooler than their counterparts around A stars, with
virtually all such disks exhibiting temperatures of $< 70$ K.  This is
consistent with the findings of \cite{wdg03} and \cite{rhe07} of a
disk population below the sensitivities of IRAS observations.

\begin{figure}[h!]
\vspace*{5.8cm}
\includegraphics{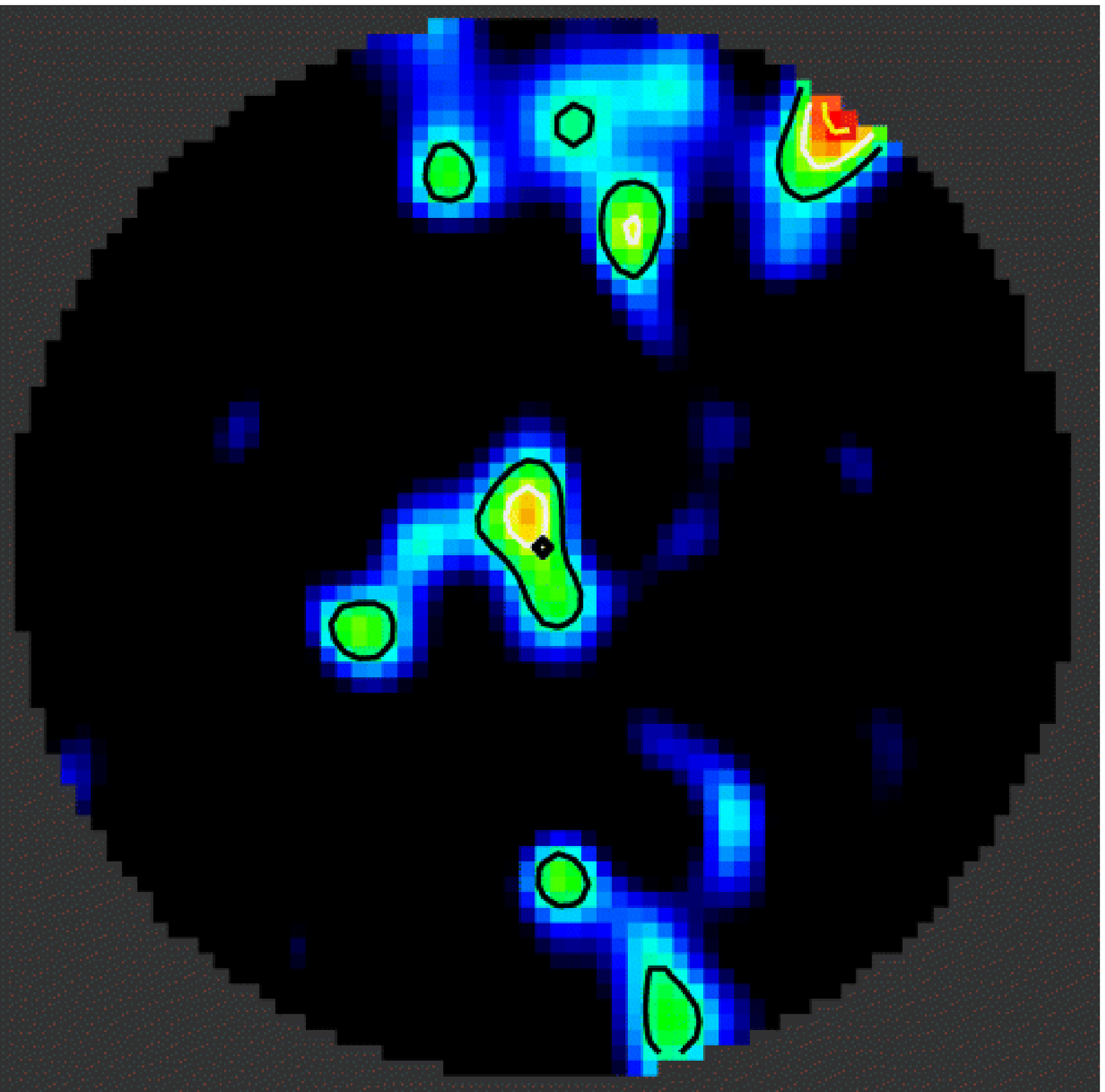}
\caption{An example of disk extraction
  \citep{gre04b} is the disk detection toward $\tau$ Ceti with SCUBA
  at 850 \micron. The rms in this map is 0.7 mJy (per smoothed 20\arcsec\
  beam).  The faintest detected peaks are 2 \mjybeam.  The disk is
  recognized by its elliptical shape centered about the star.  This
  star is nearby but the disk is in fact one of the smallest yet
  measured.  In addition, this field is unusually rich in background sources. }
\label{tauceti}
\end{figure}

To sample the cool dust, we must go to submillimeter wavelengths. The
number of unbiased surveys conducted in the submillimeter, however, is
very limited (see Table \ref{surveys}). In most cases, once the
submillimeter detections were made, only a few subsequent detailed
analysis of far-IR data showed detections \citep{zuc01}, indicating
that the dust is indeed cool. A further point illustrated in Table
\ref{surveys} is that the percentage detection rate increases sharply
with sensitivity, from 3\% in the least sensitive survey up to 10-15\%
in deep searches (and higher in the $\beta$ Pic group, perhaps because
these stars are young).  Extrapolating to still deeper surveys, we
might expect a substantially higher detection rate of about 15\%.  In
summary, there is growing evidence of a significant population of
submillimeter-bright disks with $T \le 40$ K.  The final column in
Table \ref{surveys} shows depths reached thus far (where equivalent
flux limits of other wavelengths have been corrected to 850 \micron).
The SUNS survey, with an rms sensitivity of $1 \ \sigma = 0.7$ mJy
will be {\it 40 times deeper} in flux than the only previous large
scale survey \citep{car05}.  This is the equivalent depth to the
observation of the disk around the 1 Gyr old star $\tau$ Ceti (Figure
\ref{tauceti}).

Below, we discuss the effective limits on the submillimeter survey
compared to the current generation of far-IR detectors on {\it Spitzer}.

\subsubsection{Photospheric Contributions}
For the nearby stars where photospheric emission dominates (Figure
\ref{epsiloneri}), the dust detection limit of far-IR instruments are
set by the accuracy with which a disk excess can be discriminated from
photospheric contributions.  The limits on the detectability of a disk
are then dependent as well on the accuracy of synthetic stellar
spectra, which are not well-constrained by observations at these
wavelengths.  In this case, it is not possible to improve the
detection limits, and, in the case of {\it Spitzer}, detections are
limited to $> 50$\% of the photospheric flux \citep{bei05}.  As the
$\epsilon$ Eri SED shows, the excess flux over the photosphere is
larger by a factor of 4 at 850 \micron\ compared with 70 \micron, and
this contrast becomes much greater for cooler dust.  Note that for the
very bright A0 star Vega at only 8 pc, the photosphere is only 5
mJy. Therefore, assuming SCUBA-2 calibration to 10\%, the largest
photospheric error in the survey at 850 \micron\ will be $\pm 0.5$
mJy, so the calibration error will {\it always} be less than the
confusion limit fluctuations of 0.7 mJy rms.

\subsubsection{Background Confusion Limit}

Background confusion is the absolute flux limit for any disk survey.
For {\it Spitzer}, the confusion limit is 0.7 mJy beam$^{-1}$ at 70
\micron\ (the best wavelength for debris detection), but the
photometric technique actually limits this to $\approx 2$ mJy
\citep{bei05} and often only 5 mJy is reached because of Galactic
foregrounds.  For 5 mJy at 70 \micron\ and grain emissivity of $\beta
= 0.7$, the SUNS survey is {\it intrinsically} more sensitive to dust
cooler than 30 K.  For far-IR limits of 2 mJy, single temperature
disks would be detectable down to 26 K and for a range of $\beta$ of
1.0 to 0.0, the limit lies between 26-43 K.

\subsubsection{Cirrus Confusion}

The 20\arcsec\ beam of {\it Spitzer} at 70 \micron\ and the bright and
complex cirrus background are further limiting factors to the far-IR
sensitivity.  With SCUBA-2, we lose less than 20\% of stars (i.e.,
those too far south) and importantly, with absolutely no additional
bias due the cirrus background, which is not a factor in submillimeter
observations.

\subsubsection{Extended Disks}

Submillimeter single dish observations of debris disks are constrained
by relatively poor resolution compared to optical observations. The
beamsize of the SUNSS observations will be 15\arcsec, the JCMT's
resolution at 850 \micron.  However, the potential does exist to
follow-up detections with 450 \micron\ observations with a
substantially better resolution of 7\arcsec.  Even out to a distance
of approximately 30 pc, we might expect significant extended emission based
on the largest examples with $\sim 300$ AU radii (i.e., 20\arcsec\
across) \citep{she04}, and so accurate photometry requires
fully-sampled maps to the same limiting sensitivity.  Existing surveys
are limited by the sparse pixel spacing of current instrumentation
\citep[e.g.,][]{wdg03}. SCUBA-2's Nyquist sampling will be a valuable
new feature, which is important if we are looking for relatively cool
extended dust around nearby stars.  

Our detections of new debris disks will provide some clear possible
followup observations.  Filling in the SED and thereby refining dust
temperatures and spectral indices can be carried out in the short
submillimeter and far-IR, and this prospect is imminent through the
current AKARI all-sky survey  and targeted observations with the {\it
Herschel Space Observatory} using its photometric instruments SPIRE
\citep[the Spectral and Photometric Imaging REceiver,][]{gri06} and
PACS \citep[the Photodetector Array Camera and Spectrometer,][]{pog06}.

\subsection{Mass Sensitivity}
\label{masssen}

While the SUNS survey dataset will have a uniform flux sensitivity,
the mass sensitivity of the survey is a function of three factors: the
distance of the target, $d$, the temperature(s) of the material in the
disks, $T_d$, and the opacity of the disks, $\kappa_\nu$.

\begin{equation}
M_d = \frac{F_\nu \ d^2}{\kappa_\nu \ B_\nu(T_d)}
\end{equation}

At submillimeter wavelengths, in the Rayleigh-Jeans limit, the mass of
the disk becomes a linear function of the temperature in the disk.
This makes the mass relatively straight-forward to estimate if the
temperature can be constrained.  The opacity in the disk is wavelength
dependent ($\kappa_\nu \propto \lambda^{-1}$) and difficult to measure
observationally, but the typical value adopted for debris disks at 850
\micron\ is 1.7 cm$^2$ g$^{-1}$, based on a modified blackbody with
$\beta = 1$ \citep{den00,pol94}.

\begin{figure}[h!]
\vspace*{6cm}
\includegraphics{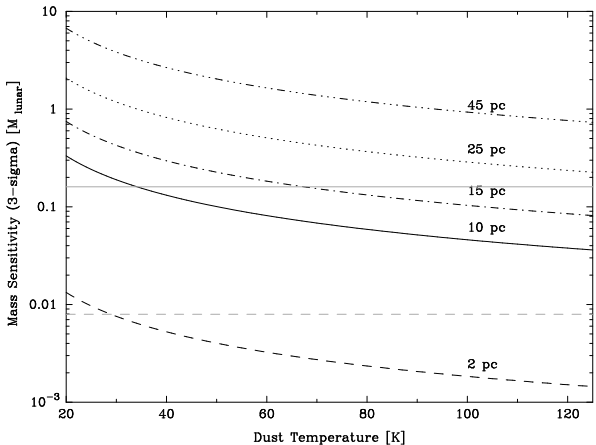}
\caption{The disk mass sensitivity (3 $\sigma$) as a
  function of dust temperature is shown for various limiting
  distances. The outer bounds of the A stars sample is roughly 40
  pc. The limits for F and G stars are $\approx 25$ pc, and
  all the M stars lie within the 10 pc boundary. The horizontal lines
  show the mass of the $\epsilon$ Eridani disk ({\it solid}) and ten
  times the mass of the Kuiper Belt ({\it dashed}).  An opacity of 1.7
  cm$^2$ g$^{-1}$ is adopted.}
\label{mass_sensitivity}
\end{figure}

Figure \ref{mass_sensitivity} shows the mass sensitivity of the SUNS
survey as a function of temperature for circumstellar dust emission at
several key distances in our survey.  The mass of the closest known
debris disk, $\epsilon$ Eridani, and the mass corresponding to ten
times that of the Kuiper Belt are also indicated.  This mass
sensitivity is for unresolved sources. For resolved sources, our mass
sensitivity will be lower.

For new disks, we won't immediately be able to constrain the disk
temperature, unless measurements (even resulting in non-detections)
have been made in the mid-IR or far-IR (e.g., through Spitzer, ISO or
IRAS).  For some cold disks ($< 50$ K), additional 450 \micron\ data
will provide some temperature constraint, but for warmer disks, both
submillimeter measurements will lie on the Rayleigh-Jeans tail of the
SED, and shorter wavelength data will be essential.

\section{Science Goals}
\label{science}

The outcome of this survey will be an order of magnitude improvement
in the number of disks known in the submillimeter.  These discoveries
will provide a significant legacy for the planet formation community
by achieving five key goals.  The SUNS survey will: 

\begin{itemize}
\item{determine unbiased statistics on the incidence of disks around
nearby stars;}
\item{constrain masses and temperatures of disks previously detected in the
far-IR (e.g., {\it by IRAS, ISO, Spitzer, AKARI, Herschel});}
\item{discover numerous disks too cold to currently detect in the mid-
  and far-IR, for which complementary data at shorter wavelengths will be
  critical to constrain temperatures and hence masses;}
\item{be the basis of source lists for future observing campaigns
using instruments such as ALMA, Herschel and {\it JWST}; and}
\item{provide limits on the presence of dust around nearby stars
that are vital to future missions such as TPF.}
\end{itemize}

Planet formation is thought to be largely complete by 10 Myr, so
studying debris disks tells us where planetesimals remain after planet
formation processes are complete. Such images may indicate locations
where planets may be located within the systems.  The majority of
debris disks we detect will be the extrasolar counterparts to the
Kuiper Belt. Thus, their temperatures, which will be determined through
complementary far-IR data, tells us the sizes of the planetary systems
which are clearing the inner regions of dust and planetesimals.
Submillimeter fluxes also provide the most reliable method for
measuring the mass of the disks \citep{zuc01}.

We can thus address questions of what affects the extent of a
planetary system and the mass of its planetesimal component.  For
example, do more massive stars form the most massive planetary systems?
Are the disks truncated uniformly with spectral type indicating that a
universal process is at play, such as close stellar encounters in the
cluster in which the star formed, or photo-evaporation of the outer
disk edge by a nearby massive star in such a cluster.  It already
appears that more massive stars host more massive disks \citep{gw03}
that are more readily detected \citep{hab01}.  That said, a comparison
of the disk masses around members of the nearby young TW Hydrae
Association suggests no correlation between disk mass and spectral
type for these reasonably coeval disks \citep{mat07}.  In addition,
comparisons of disk masses around low mass stars of comparable age,
i.e., AU Mic \citep{liu04} and TWA 7 \citep{mat07}, reveal a
difference of a factor of 10.  Improved statistics are critical to
reveal whether a genuine trend exists. Moreover, information about
disk radii is scarce at present -- the only reliable determinations
are for the few disks which have submillimeter detections
\citep{she04}, and these comprise a significantly biased dataset.

With the SUNSS data, it will be possible to test for planetary system
evolution.  Based on our understanding of the Solar System's
evolution, it is reasonable to expect that both the radius and mass of
the belts will change with time.  Within our own Solar System, the
outer planets may have migrated from the locations where they formed
\citep{mal95}, a scenario used to explain structure in some debris
disks \citep{wya03}.  For example, Uranus and Neptune may have formed
in between Jupiter and Saturn until their orbits became unstable at a
few hundred Myr \citep{tho99}.  The radius of the Kuiper Belt may have
also been significantly different in the past.  The period of late
heavy bombardment in the inner Solar System and other evidence, such
as the fact that the Kuiper Belt must have been more massive in the
past \citep{mor04}, also indicates that the dust flux from the Kuiper
Belt must have varied significantly with time. Thus, understanding the
evolutionary changes in extrasolar systems will not only help place
the evolution of the Solar System in context, but could also help to
illuminate Earth's own origins.

The survey will test the applicability of the Solar System itself.
One model for the origin of debris disks proposes that they flare into
detectability when a system matures to the extent that a Pluto-sized
object can form in its planetesimal belt \citep{ken04}.  The new
planet then perturbs the belt out of which it formed, initiating the
collisional cascade that produces the copious amounts of dust that we
see for a short amount of time.  Since such Pluto-analogies may take
longer to form further from the star, it can take up to 1 Gyr for
those outer disks to become detectable. This model would mean that we
could expect to detect disks at large distances around stars at late
times \citep{rie05}.  Short bursts can also be produced when
lunar-sized embryos are ejected from the inner planetary system.  Such
processes may explain the presence of dust around old stars, as well
as the apparently episodic nature of the debris disk phenomenon
\citep{rie05}.  Such disks would be cool and so may only be detectable
at submillimeter wavelengths.  Indeed it can be shown that for several
disks the dust we are seeing must be transient \citep{su05,wya07}.  An
alternative view is that disks may be quasi-static in radii.  These
models say that dust mass (but not radius) does evolve with time,
decaying through the collisional erosion of the belts \citep{dom03},
and such models can successfully explain the currently available
statistics (Wyatt et al., submitted).  Episodic periods of high dust
mass are then provoked through the destruction of large planetesimals
\citep{wd02} or a recent stellar flyby \citep{kal01}.  Our survey will
confront these models with hard statistics.

The unbiased nature of the SUNS survey means that it will be possible to
correlate disk parameters with other planet formation indicators. For
example, we can assess whether the presence of a giant planet orbiting
near the star, or a distant planetary companion detected with adaptive
optics, affects the mass or size of an outer disk.  Such planets may
form and evolve independently from the planetesimals in the system's
outer reaches.  However, evolutionary processes which lead to planets
migrating inward (e.g., massive gas disk, fast core growth) may be
related to evolutionary processes further out, either promoting or
inhibiting the presence of cold disks.

\begin{figure}[h!]
\vspace*{15cm}
\includegraphics{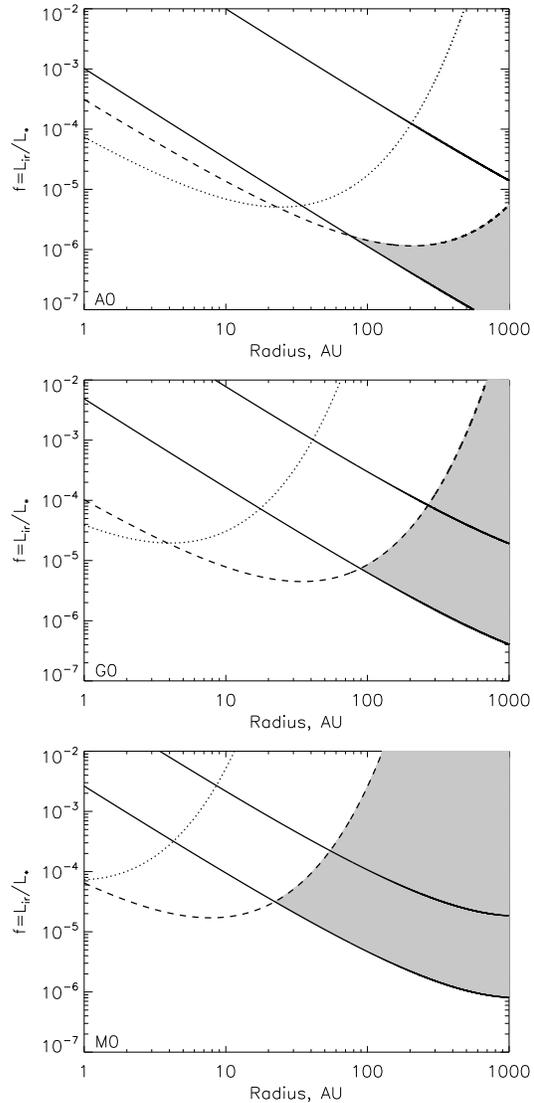}
\caption{These plots show the parameter space in
  which the SCUBA-2 survey is uniquely sensitive ({\it shaded area}).
  In each case, this is the location of cold dust. The different lines
  show the limits of Spitzer 24 \micron\ sensitivity ({\it dotted
  line}), assuming a limit of $F_{\rm disk} / F_* = 0.1$ and Spitzer
  70 \micron\ sensitivity ({\it dashed line}), assuming a limit of
  $F_{\rm disk} / F_* = 0.55$ based on \cite{su07}. The solid lines
  are the SCUBA-2 3 $\sigma$ limits for the closest and furthest stars
  in the samples, assumed to be 2.6-45.2 pc for A0, 3.6-24.8 pc for
  G0, and 1.8-8.6 pc for M0.  For A0 stars, this space is relatively
  narrow (shown by the shaded region). For M stars, however, the
  region is much more extensive and includes significantly smaller
  disks below the detection limits of Spitzer at 70 \micron.  This
  figure does not account for the fact that at these distances the
  largest disks are going to be significantly resolved, and therefore,
  we won't reach the 2 mJy limit, i.e., the SCUBA-2 limits are
  optimistic for disks larger than the beam size.}
\label{parameterspace}
\end{figure}

These are issues which can only be resolved with a large unbiased
survey so that the variations of disk frequency with spectral type,
binarity, and age can be untangled. The submillimeter is the ideal
waveband in which to do this, since it is not limited to warm (40-100
K) disks (and so small planetary systems). The incidence and evolution
of mass or radius of cold ($< 40$ K) disks is completely unconstrained
at present.  Using the submillimeter also means that an estimate for
the disk mass is obtained immediately, with more accurate limits on
temperature and morphology possible once complementary data are
obtained at shorter wavelengths.  As Figure \ref{parameterspace}
shows, the SUNS survey will be particularly sensitive to disks around
M stars, since the circumstellar dust is cold even at relatively close
proximity ($< 100$ AU) to the host stars. 

For the nearest disks, the SCUBA-2 observations will result in
resolved images.  During its lifetime, SCUBA imaging at 850 \micron\
imaged seven of the fourteen resolved disks
\citep{hol98,gre98,she04,gre04b,wya05,wil04}.  Images allow the inner
and outer radii of the disk to be determined directly, thereby
confirming inferences about inner cavities. The morphology
of disks can thereby reveal the presence of unseen planets
\citep[e.g.,][]{wya03}.  The orbits, masses and even the evolutionary
histories of planets have been constrained in this way. Such images
will thus be an extremely powerful tool for telling us about a region
of parameter space unreachable with other techniques: planets
perturbing debris disks are both small (Neptune size) and at large
distances (tens of AU).

For disks that are not resolved, the information returned by SCUBA-2
will provide critical information, since it can confirm that the
emission is star-centered.  Associations of excesses with stars are
not always reliable from far-IR data alone because of the relatively
large beamsize \citep[e.g.,][]{jay02}.  For unresolved sources, SED
fitting and radius estimation will indicate those that could be
resolved with future observatories, at a range of wavelengths from
optical to millimeter.  Of particular importance will be ALMA
follow-up, with our survey giving prior knowledge of the submillimeter
disk flux and good indication of angular size. These parameters can
only be guessed at from far-IR data alone, so with these data far less
ALMA time will be spent observing sources which are ultimately too
faint to detect, or performed using non-optimal array configurations.

\cite{wya07} modeled the population of disks around A stars, fitting
to the Spitzer statistics of \cite{rie05} and \cite{su07}, yielding
predictions for the outcome of the A star component of SUNSS,
excluding the cold disks which would not have been detected by
Spitzer.  They predict that 17 (of 100 stars) would have detected
disks at $> 2$ mJy and that $> 5$ would be resolvable in 450 \micron\
imaging.  This study also concluded that the SUNSS would be vital for
constraining the distribution of disk radii and for determining the
unknown population of disks that are too cold to detect with Spitzer
(not included in the 17\% of expected disks).  Assuming the other
spectral types have a similar submillimeter detectability, we could
justify a prediction for the survey of 85 detections and 25 resolved
disks based on the model of the A star population.  A cold disk
population not currently sampled by the Spitzer statistics would
result in an elevated number of disk detections. We have estimated
this value as approximately 10\% (50 disks), but deducing the true
incidence of cold disks is in fact one of the principle drivers for
this survey.

\section{Target Details}
\label{targets}

\subsection{The Sample}

The survey will cover 500 nearest stellar systems -- 100 primaries of
each of the spectral types A, F, G, K, and M observable from the
JCMT. Subgiants are included for spectral types A, F and G, while only
dwarfs are included in the K and M targets. The aim is to obtain
samples that are statistically robust and can be inter-compared, while
keeping the survey completely unbiased with regard to choice of
star. This is a unique feature of our survey; no star will be rejected
because of its intrinsic properties.  The nature of the mass function
of stars means that the five sub-samples cover different volumes. The
distance limits extend out to 45, 25, 22, 16.5 and 8.6 pc for A, F, G,
K and M stars respectively. The uncertainties on the parallax for
these stars are $\pm 1$ pc for A, F and G stars and $\pm 0.1$ pc for K
and M stars. These volume explored by this survey are similar to those
being explored with {\it Spitzer} and earlier with {\it ISO}, but the
SCUBA-2 survey will be the first unbiased study.  Further details on
the selection of targets will be presented in a forthcoming paper
(Phillips et al., in preparation).

The allocation of 390 hours includes 330 hours during the first two
years of SCUBA-2 operations.  Of this time, 270 hours has been
allocated to weather band 3 ($0.08 < \tau_{\rm 225 GHz} < 0.12$) and
60 hours is available within band 2 ($0.05 < \tau_{\rm 225 GHz} <
0.08$) weather.  (The lowest-elevation stars are prioritised for the
better conditions.)  The number of targets that need close observing
constraints (having low elevation and thus priority for upper band 2
conditions) is only 10\% of the total number of targets. The remaining
60 hours is allocated beyond the first two years of operation in band
2 conditions; some followup may be possible in this period, although
completion of the 850 \micron\ survey will have priority.

\begin{figure}[h!]
\vspace*{6.8cm}
\includegraphics{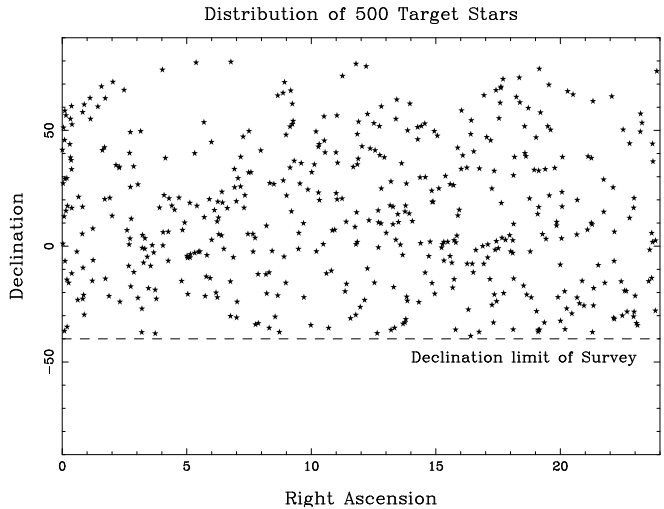}
\caption{Distribution of target sources in
  the DDS.  The targets are all-sky. The declination limits are $-40 <
  \delta < +80$. }
\label{distribution}
\end{figure}

Figure \ref{distribution} shows the targets are distributed across the
sky; the furthest distance is $45\pm1$ pc, and there is no clustering
of stars towards the Galactic Plane.  Observations at any time of the
year will have suitable targets. The declination limits of $-40$\degr\
to $+80$\degr\ ensure that targets rise to at least 30\degr\
elevation.

We will image the total sample of 500 stars down to our adopted
JCMT confusion limit of 0.7 mJy at 850 \micron.  In borderline band 2/3
conditions ($\tau_{\rm 225 GHz} = 0.08$) at airmass 1.15 (30\degr\
from zenith), this sensitivity requires an average of 45 minutes
observing per star. The priority will be for 850 \micron\ observations
with 450 \micron\ imaging a bonus (giving first hints of disk
sub-structure suitable for follow-up). In general, 450 \micron\
observations will form a follow-up PI-driven project once the
population of suitable detected disks is known.  The practical
strategy is to start with the nearest targets and work outwards, while
maintaining balance among the spectral types. Thus the best resolved
disks will be observed first, and these will have the highest priority for
follow-up imaging.

Figure \ref{mass_sensitivity} shows the variation of mass sensitivity
for disks of different temperatures and the limitations for several
key distances in the survey, including the inner bound (2 pc)
and the outer bound (45 pc), as well as the outer bounds of the
M dwarf (8.6 pc) and G and F stars (22 pc and 25 pc).  This figure
shows that, for unresolved disks, our sensitivity is sufficient to
detect masses comparable to that of $\epsilon$ Eridani (0.002
$M_\oplus$ = 0.16 $M_{\rm lunar}$) at most temperatures within 10 pc.
The mass of $\epsilon$ Eridani lies in the mid-range of dust masses
for the handful of detected Sun-like systems \citep{gre04}. The
detectable mass is of course modified by dust location, stellar
heating, and resolution, but we note that 15 pc is a useful reference
point as it is the far end of planet search distances for TPF.  At
that distance, we will detect all unresolved systems with temperatures
exceeding 70 K.
Such limits mean that non-detections result in constraints on
dust mass in these systems of typically 200 times the dust content of
the solar system, but down to only approximately four times this for
our nearest target, GJ 699, at 1.8 pc.

\subsection{Ancillary Targets}

In addition to the 500 primary targets of the survey, the large FOV of
SCUBA-2 will enable us to image companions in multiple systems, and in
some fields, to include other nearby stars where they are present.  In
our target list, at least 240 stars are members of multiple systems of
at least two members. This high fraction of multiple systems raises
the total number of stars surveyed to well over 750.  Very few of
these systems require more than one SCUBA-2 sub-array, since even wide
binaries up to 1000 AU in separation will have an angular separation
of $< 2$\arcmin.  In the case of binaries, their inclusion in the
survey is based on the spectral type of the primary; exempting the
secondary from the initial sample ensures that no bias is introduced
due to the probabilities of disks in binary systems or the weighting
of secondaries toward lower masses.

\newpage
\subsection{Subset Populations}

The sample-size for each of the spectral types was chosen so that we
could statistically distinguish between detection rates of 5, 10, 25
and 50\% when the primary dataset is divided into the following
sub-categories.

\begin{itemize}
\item{Stellar type, with 100 stars each of A, F, G, K and M. (Properties
  here and subsequently are from the NASA NStars database at
  http://nstars.arc.nasa.gov).}
\item{Stellar age, with $\sim 150$ stars younger than 1 Gyr and $\sim
  350$ stars aged 1-10 Gyr. The division corresponds approximately to
  the end of the heavy bombardment phase of the Earth, and this age
  split can readily be applied from the decline in X-ray activity, as
  measured by the ROSAT all-sky survey \citep{gai98}. The 150:350
  division arises from the proportions within each stellar type
  falling with this age bracket. This is $\sim$100\% of A-stars (1 Gyr
  on the main-sequence), 20\% of F-stars (lasting 5 Gyr) and 10\% of
  G, K and M stars (limited by the Galactic Disk age of 10 Gyr).}
\item{Stellar multiplicity, with $\approx 50$\%  of target stars having another star as
  a companion. }
\item{Presence of a planetary system, with $\sim 20$ examples within our volume
  known at the present time from radial velocity searches.}
\end{itemize}

\subsection{Statistics}

The overall choice of sample size is driven by the aim of
differentiating detection rates among the listed sub-groups. The key
is to have a sample size large enough that the smallest sub-group with
the lowest detection rate can be distinguished from rates in other
sub-groups. The detection rate for X disks among Y stars is defined as
$Z = (X +/- \sqrt{X}) / Y$ as the uncertainty is Poissonian for small
counts. In the simplest interpretation, we accept rates as
differing if the upper bound of one Z value is below the lower bound
of the comparison Z value. In practice this means the survey size is
driven by the smallest samples, that are one-tenth of the comparison
group. The two cases are the planetary systems ($\sim$ 1 in 10
Solar-type systems) and young Sun-like stars (20 among 200 of the G
and K stars). Here we can distinguish observed detection rates $Z$ for
any pair of intrinsic detection rates among 10\%, 25\% and 50\%. The
formal minimum required to do so is 163 stars, which is just met by
our smallest group: the approximately 180 stars of late-F to mid-K
potential planet-hosts.

Detection rates do not need to be compared within every spectral type.
For example, the effects of age on debris will be investigated for
stars of Sun-like mass, i.e., types both G and K, rather than G and K
separately. In fact, this approach is required to allow us to
establish robust distinctions (i.e., at the 2-3 $\sigma$ level)
between detection rates as a function of spectral type and age, due to
the practical limits on sample size. For instance, with 200 stars per
sample, a 15\% difference in detection rate can be identified at the 3
$\sigma$ level.

Results from the far-IR surveys with {\it IRAS}, {\it ISO} and {\it
Spitzer} show that these rates actually occur amongst sub-groups of
stars. For example, 10\% is the global rate for Sun-like stars
\citep{gw03}, 25\% is the rate among Sun-like stars with known giant
planets \citep{bei05}, and 50\% is the rate among more luminous A
stars \citep{gw03,rie05}. It is useful to note that 10\% of disks
exist even at the limited {\it IRAS} sensitivity, and so this a
minimum frequency. However, we can also distinguish a 5\% disk
occurrence should it occur among one of the less well-studied
sub-groups. In the smallest sub-groups of only 20 stars, we have
evidence already for detection rates of 20-25\% \citep{gw03,bei05}.

\section{Complementary Data}
\label{complementary}

To make the most of the JCMT Legacy survey dataset, it is essential
that we ensure that the targets are selected based on accurate
spectral classification.  Although the targets are all very near the
Sun, this proximity does not guarantee that the optical spectrum of
each is well-characterized.  This is critical to exclude K-giants and
brown dwarfs, as well as to ensure that the sub-categories of the
targets are accurately populated.  As part of the ongoing RECONS
survey of nearby stars, \cite{hen06} have recently published spectral
classifications for most of the nearby stars on our target list.  For
those which have not yet been observed spectroscopically, we have
undertaken a small optical survey. We are using the 1.2 m telescope at
the Dominion Astrophysical Observatory to obtain the bulk of these
spectra with a resolution $R = \delta \lambda / \lambda \approx 3000$.

The SUNS survey is complementary to {\it Spitzer} programs searching
for warmer debris disks at 24, 70 and 160 microns. Synergy with {\it
Spitzer} data is important to construct the spectral energy
distributions of the full disk population, even where {\it Spitzer}
has failed to detect a disk in our targets. Due to the timescales of
the missions, most far-IR complementary data for our survey targets
are likely to come from AKARI and the Herschel Space Observatory.
AKARI is currently doing an all-sky far-IR survey. 
A targeted Herschel debris disk survey is planned using guaranteed
time on the Herschel photometric instruments PACS and SPIRE. In
addition, other photometric observations are likely to be proposed
during open time.

\section{Data Products}
\label{dataproducts}

The data products of the survey will include an archive of 500 single
fields centered on the 500 target stars, publication quality images of
all the detected disks, and a catalogue of the measured fluxes and
flux limits at 850 \micron\ and 450 \micron.  In addition, plots of
the fitted SEDs for each disk and star combination will be made using
optical and IR data combined with the submillimeter fluxes.

It will be possible to extract systematically parameters from the data
\citep[see][]{she04}. These will also be included in the Legacy
database of the survey, and include dust temperatures,
masses, spectral indices, and characteristic disk radii (either
inferred from temperature and spectral index, or measured where the
disks are resolved).  Systematic modeling will also yield masses of
colliding bodies within the debris belts (up to specified sizes), and
estimates of mass, location and orbital direction for perturbing
planets (where the disk structure is well-resolved).

We have already developed additional tools for measuring disk sizes
and fitting spectral energy distributions and will make these publicly
available. For examples of development of public databases that we
have already made available\footnote{see
http://www.roe.ac.uk/ukatc/research/topics/dust/indentification.html}. Further,
our {\it Spitzer} contributors have developed tools using HIRES
techniques, specifically for resolution-enhanced mapping of debris
disks. These tools will be used in a unified analysis of the
SCUBA-2/{\it Spitzer} database.  Herschel data will be incorporated to
our analysis where possible.

\subsection{Ancillary Data Products}

Some important ancillary data products will emerge from the survey.
These will include a catalogue of fluxes and positions of background
sources not associated with the star, the majority of which will be
distant, dusty galaxies of the kind discovered by \cite{sma97}, and a
list of fluxes for nearby stars where the photosphere is detected but
there is no debris disk.

The extragalactic catalogue has value in providing a large database of
submillimeter galaxies (SMGs) that can be followed up with adaptive optics to
measure optical/IR properties at very high spatial resolution. A
unique aspect of these target fields is that the survey stars
themselves will be available for guiding, and they are technically
ideal, with magnitudes down to about $m_V = 12$ and offsets of up to
about 5 arcminutes. Further, the fields add up to 7 square degrees of
sky observed to the JCMT's 850 \micron\ confusion limit but split 500
ways rather than in a few fields. Thus the sources counts provide a
test for cosmological variance, in comparison to the SCUBA-2 cosmology
survey.  Each 49 square arcminute survey field is well-matched to the
field of view of the KMOS near-IR facility on ESO's 8-m VLT
\citep{sha04}, allowing up to 24 SMGs to be targeted spectroscopically
using deployable integral field units, which promises to yield the
least-biased spectroscopic redshift distribution to date. 

The stellar photospheric data are valuable because the long-wavelength
end of the Rayleigh-Jeans tail has never been measured. As an example,
the A0 star Vega at 8 pc has a photosphere of about 5 mJy at 850
\micron, and so the equivalent for a non-debris star can be detected
with our survey at the $7\sigma$ level. The science value is in
observing the contribution of the photosphere to the SED of the star,
a quantity that is poorly measured in the far-IR, especially for
cooler stars (K and M) with deep molecular absorption bands. Measuring
signals below the expected levels would show that the photosphere does
not simply extrapolate as a blackbody, constraining stellar atmosphere
models. At present, about 5\% of stars are known to have far-IR
signals $3 \ \sigma$ or more below the photospheric prediction
\citep[e.g.,][]{hab01}.

\section{Summary}
\label{summary}

Many stars are surrounded by dusty cold debris disks.  These are fed
by asteroids and comets orbiting the stars. Studying the location,
mass and morphology of these disks provides crucial information about
the outcome of planet formation in these systems.  For the fourteen
disks which have been resolved at present, observed structures have
even been used to suggest the location of unseen planets. Hundreds
more stars have had their disks characterised from their SEDs showing
that these disks are the extrasolar equivalents of the Kuiper and
asteroid belts in the Solar System.

In this Legacy survey, we will use 390 hours of SCUBA-2 time on the
JCMT to observe 500 nearby main-sequence stars to search for debris
disk signatures. This survey will be the first unbiased one since
IRAS, as previous far-IR surveys have had to omit many
stars. The crucial value of the submillimeter is that the stellar
photospheric signal is irrelevant and so any star can be examined. The
output of our survey will be robust statistics on the incidence of
debris disks plus discovery of the underlying causes (in terms of the
stellar environment and history). The nearer systems may also be
resolved, contributing to planetary detection and planning for missions
such as TPF/Darwin.

The data products will be unique, comprising deep and uniform searches
for debris without any bias towards particular stellar
properties. This has never been done at any wavelength, and
particularly not in the submillimeter where a new cold population of
disks is barely explored.  The SUNSS will exceed the modest, unbiased
G-dwarf SCUBA survey \citep{gw03} by forty-fold in stellar numbers
while being substantially deeper. The SCUBA-2 sensitivity will
approach the Kuiper belt dust level for the closest Solar analogues; a
disk around these targets could actually be detected in our survey
before the equivalent has been mapped around the Sun. The survey can
never be done better until large far-IR telescopes fly in space
-- resolving the disk spatially from the stellar photosphere -- a
prospect considerably downstream of JWST.

The Science Legacy lies in answers to the five key outcomes:
\begin{itemize}
\item{determining unbiased statistics on the incidence of disks around nearby stars;}
\item{constraining masses and temperatures of disks previously
  detected in the mid- and far-IR (e.g., in {\it Spitzer} surveys);}
\item{discovering numerous disks too cold to detect in the far-IR;}
\item{being the basis of source lists for future observing campaigns
  using, e.g.,  ALMA and {\it JWST}; and}
\item{providing limits on the presence of dust that are vital to future missions such as Darwin/TPF.}
\end{itemize}

With these answers in hand, we will be able to understand for the
first time the relation of debris disks -- tracing planetesimals up to
tens of km across in orbits at tens of AU -- to the inner planetary
systems detected by other methods. The results, especially when
combined with shorter wavelength data to constrain temperature and
mass, will test models of planet formation spanning across the scale
of our Solar System (from inside Mercury's orbit to beyond
Neptune's). The images of disks will be followed in the next decade by
high-resolution imaging that may indicate perturbing planets, even
following their orbital perturbations in real time. The results will
be vital for the detection of extrasolar Earths with coronagraphs.

\acknowledgements

\end{document}